\documentclass[aps, twocolumn,floats, showpacs,floatfix]{revtex4}
\usepackage{amsmath,amssymb,graphicx}
\usepackage{epsfig}
\usepackage{graphicx}
\usepackage{enumerate}

\newcommand{\BigO}[1]{\ensuremath{\operatorname{O}\bigl(#1\bigr)}}
\newcommand{\beq}{\begin{equation}}
\newcommand{\eeq}{\end{equation}}

\newcommand{\bea}{\begin{eqnarray}}
\newcommand{\eea}{\end{eqnarray}}

\newcommand {\ket}[1]{|\,{#1}\,\rangle}
\newcommand {\bra}[1]{\langle\,{#1}\,|}

\begin{document}
\title{
%Non-adiabatic transitions in a linearly driven two-level system mediated by a discretized continuum}
Landau-Zener transitions mediated by an environment: population transfer and energy dissipation}
% energy fluctuations}
%Nonadiabatic transtions in level crossing
\author{Amro Dodin$^1$, Savannah Garmon$^2$, Lena Simine$^1$, Dvira Segal$^1$}
\affiliation{$^1$Chemical Physics Theory Group, Department of Chemistry, University of Toronto,
80 Saint George St. Toronto, Ontario, Canada M5S 3H6}
\affiliation{$^2$Institute of Industrial Science, University of Tokyo, Komaba 4-6-1, Meguro, Tokyo 153-8505, Japan}

\date{\today}
\begin{abstract}
We study Landau-Zener transitions between two states with
the addition of a shared discretized continuum.
The continuum allows for population decay from the initial state as well as indirect transitions between
the two states.
The probability of nonadiabatic transition in this multichannel model 
preserves the standard Landau-Zener functional form
except for a shift in the usual exponential factor,
reflecting population transfer into the continuum.
We provide an intuitive explanation for this behavior assuming independent individual transitions between pairs of states.
In contrast, the probability of survival in the ground state at long time shows a novel, 
non-monotonic, functional form,  with an oscillatory behavior in the sweep rate at low sweep rate values. 
We contrast the behavior of this open-multistate model 
to other generalized  Landau-Zener models incorporating an environment:
the stochastic Landau-Zener model and the dissipative case, where energy dissipation 
and thermal excitations affect the adiabatic region.
Finally, we present evidence that the continuum of states may act to shield the two-state Landau-Zener
transition probability from the effect of noise. 
\end{abstract}

\maketitle
%--------------------------------------------------------------

\section{Introduction}

%Historical Context
Systems that may be modeled by avoided level crossings are ubiquitous in nature and in artificial mesoscopic 
systems. In 1932, Zener \cite{Zen}, Landau \cite{Land}, Stueckelberg \cite{Stueck}, and Majorana \cite{Maj} 
separately derived an expression describing the probability of nonadiabatic transitions 
at avoided level crossings, based on semi-classical modeling.
The result, typically referred to as the ``Landau-Zener" (LZ) formula, has been applied to describe transition probabilities
in the context of chemical reactions \cite{Reaction}, production of cold molecules \cite{ColdRev}, 
quantum electrodynamic circuits \cite{QED1, QED2},
spin-flip in nanomagnets \cite{nanmag},
Bose-Einstein condensates in optical lattices \cite{BEcond}, 
%SSS
doublon-hole production in a Mott insulator \cite{Oka},
directed quantum transport in bipartite lattices \cite{bipartite}, 
and adiabatic computing \cite{QAdi, AdComp}.
Furthermore, the LZ model has been used to design the Landau-Zener-Stueckelberg spectroscopy 
technique \cite{LZS}.

%Extension of the Model
The original Landau-Zener problem is restricted to two coupled diabatic levels 
%coupled with a ixed tunneling element
%$|0\rangle$ and $|1\rangle$% 
with an energy gap that is linearly modified in time at a 
constant rate. 
Given a certain initial state, the ground diabatic state, the quantities of interest are the 
population probabilities for both the ground and the excited diabatic states at infinitely long time, far from the crossing region.
This minimal LZ situation has been revisited many times 
and extended to a broader class of systems; in particular, the multistate Landau-Zener problem, a 
representative of time-dependent Hamiltonians, has been studied analytically and numerically in Refs. 
\cite{DemOsh,Brundobler,Nogo,Sym,Sinitsyn,BT,NBT,GBT1,GBT2,SpinBath,Akulin,Moyer,Band}. % ADD
%
%The LZ problem with the continuum, as depicted in  Fig. \ref{Fig1}(a),
%can be regarded as a multistate LZ model.
%Indeed many variants of the multistate case have been  explored in the literature \cite{BE,Nogo,Sym,BT,DemOsh},
%as they offer a simple setting for studying the dynamics of a driven system.
In such models it has generally been argued that the probability to remain in the 
initial diabatic state (in other words, to make a nonadiabatic transition)
follows a semiclassical behavior: It decreases exponentially with the number of 
%SSS
avoided crossings $N$, 
$P_{na}\sim e^{-\beta N}$, where $\beta$ characterizes the transition probability at an individual crossing. This form is valid as long as the energy of the initial state
follows a linear dependence in time \cite{Sinitsyn}. 

%==============================

\begin{figure}[htbp]
\includegraphics[scale=0.4]{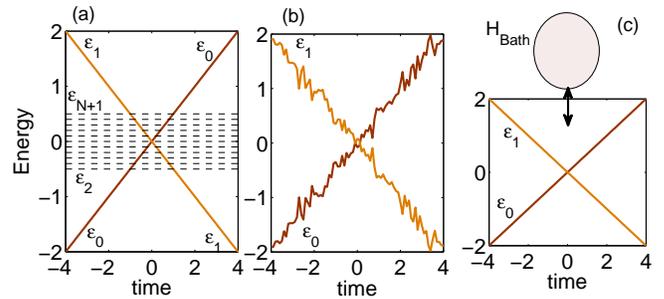}
\caption{
Schemes of the different models studied in this work.
(a) Open-multistate Landau-Zener problem.
The diabatic energies of the original LZ problem are presented by full diagonal lines.
The dashed lines represent stationary states of a dense continuum. 
Parameters correspond to the sweep velocity $\alpha=1$, tunneling element $\Delta=0.1$ 
and the band cutoff $D=0.5$.
(b) Stochastic-Markov LZ model. 
The energy difference between the diabatic states fluctuates around the original LZ value. 
%Here the  variance $\kappa^2=XXX$.
(c) Dissipative LZ model with the diabats coupled to a (finite-temperature) 
harmonic heat bath.
}\label{Fig1}
\end{figure}
%==============================

% environment
%Understanding the role of an environment on the LZ transition is important since in 
%physical situations a quantum system is never truly isolated from an environment, possibly leading to 
%
%SSS
However, in physical situations a quantum system is never truly isolated from its environment, which can have the effect of inducing
phase decoherence, population relaxation, energy dissipation, and bath-induced transitions.
This problem has been studied extensively in different regimes of sweeping speed, 
temperature, and system-environment coupling strength,
using an array of perturbative approaches;
see for example the early study of Tsukada \cite{Tsukada}
and other comprehensive works \cite{Gefen,Ao,Stern, KayanumaPRB,Sun,spin,Ziman,Band}. 
While the transition probability can be obtained exactly-analytically 
when the diabats are coupled to a zero-temperature bath \cite{HanggiT0}, 
at nonzero temperatures, 
numerically-exact simulations %of the LZ dynamics in the presence of a finite temperature harmonic bath
have been reported in Refs. \cite{Thorwart,LeHur}, %utilizing a harmonic bath,
revealing a rich dynamics. Specifically, it has been 
shown that at a nonzero temperature the (harmonic) heat bath is responsible for 
transition probabilities that are 
non-monotonic in the sweep velocity, the result of a nontrivial competition between the driving (sweep) and 
bath-induced excitation and relaxation processes \cite{Thorwart}. 
%The LZ model thus offers a minimal setup for exploring the role of coherence, and thermal effects
%in affecting driven quantum dynamics and reaction dynamics.

%Brundobler and Elser published a formula for the scattering matrix element between states in the limit of independent transitions \cite{BE}, while Sinitsyn has 
%formulated the no-go theorem \cite{Nogo} and symmetry conjecture \cite{Sym} for multistate LZ models.

% HEre
%
%SSS
Our primary objective in this paper is to study the nonadiabatic transition probabilities of
the {\it open-multistate LZ model}, consisting of two driven states (as in the original LZ model)
coupled to a finite-band discretized continuum consisting of stationary states.  
The avoided crossing between the driven states occurs directly at the center of the finite band, as shown in
%In this paper our main objective is to study nonadiabatic 
%transition probabilities at avoided level crossings in 
%an open-multistate LZ model, with the driven two states coupled to a finite-band 
%discretized continuum made of stationary states, see 
 Fig. \ref{Fig1}(a). The continuum allows for indirect transitions between the two main diabats, mediated 
by population transfer to the continuum. This setup can be used to model, 
for example, multichannel reactions with 
intermediates and competing products \cite{Tully1,Tully2}. In the context of mesoscopic devices, the model
can serve to describe
all-electrical population transfer processes between spatially 
separated quantum dots coupled indirectly via a chain of intermediating dots 
\cite{CTAP,Kohler} or through a central metal \cite{Gurvitz}.
%SSS
One could also consider quantum memory preservation in a spin quit
interacting with its environment, or a similar system.  Here environmental
processes may induce in an unintended spin flip, resulting in memory loss.  Understanding
these processes  should make it possible to better control them.

%Our goal here is to identify signatures of population decay and  indirect transfer mechanisms between the LZ diabats, emerging from competing pathways. 
%The signatures of population relaxation and indirect transitions between the diabats 
%should be contrasted to, and separated from,
% the fingerprints of energy dissipation and thermal excitations, the result of coupling the diabatic 
%states to a heat bath. 

We note that the LZ model with the discretized continuum, as depicted in  Fig. \ref{Fig1}(a), 
is an example of a multistate LZ model.
Using numerical simulations we show below that while the nonadiabatic transition probability 
$P_{na}$ follows an exponential decaying form
%SSS
(similar to other multistate LZ models appearing in the literature as discussed above) 
the ground state survival probability in our model displays nontrivial and non-monotonic 
features, revealing indirect pathways between the two diabats. We explain these signatures 
and further contrast them to the fingerprints of energy dissipation and thermal excitation processes.
We do so by constructing 
several elementary models as shown in Fig. \ref{Fig1}:  
(a) The open-multistate LZ model, as discussed above, with the two driven states coupled to a 
finite-band discrete stationary continuum. 
%This continuum allows for  population relaxation and indirect transtions
%between the LZ diabats.
%Physically, this simple single-body model represents
%nonadiabatic transtions at avoided level crossings in the presence of competting pathways.
(b) The stochastic (infinite temperature) model, where the isolated LZ Hamiltonian suffers
from noise, responsible for energy fluctuations \cite{Kayanuma-stoA}. 
(c) The dissipative finite temperature LZ many-body model, 
where the two driven states are bilinearly coupled to a harmonic heat bath. 
%driving, relaxation and dissipation.

% Specifically, we are intereseted in the signatures
%signatures of competing

The paper is organized as follows. 
In Sec. \ref{MODEL} we describe variants of the LZ model that allow us to 
pinpoint how the original LZ problem is modified by
its coupling to a stationary continuum or a heat bath.
%responsible for population transfer, as opposed to
%the role of a heat bath, allowing for energy dissipation and thermal excitations.
Numerical results are included in Sec. \ref{RESULT} along with detailed analysis. 
In Sec. \ref{CONC} we discuss and conclude our observations.
For simplicity, we set $\hbar= 1$ and $k_B=1$ throughout this paper.

%-------------------------------------------------------------

\section{Models}
\label{MODEL}

\subsection{Isolated Landau-Zener model}
\label{model1}

The original  LZ model includes two diabatic states $|0\rangle$ and $|1\rangle$ with a fixed 
tunneling matrix element $\Delta$. The energies  of these states are
modified linearly in time, 
$\epsilon_{0}=\alpha t/2$, $\epsilon_{1}=-\alpha t/2$ at a sweep rate 
$\alpha$ (dimension Energy$^{2}$), 
\bea
H_{LZ}^{iso}(t)&=&\epsilon_0(t) \ket0 \bra0 +\epsilon_1(t) \ket1 \bra1
\nonumber \\ &+& \Delta \ket0 \bra1 +{\Delta} \ket1 \bra0.
\label{eq:HLZ}
\eea
Since the two states are isolated (iso) 
from an environment, we refer to this as the ``closed  LZ model."
At large negative time, before approaching
the avoided crossing, the system is prepared in the ground state corresponding to $\ket0$. 
We define the ``survival probability" $P_s$ as the probability for the system to end up at long time
in the ground state (thus cross diabats).  Meanwhile $P_{na}$ describes 
the probability to stay on the same state $|0\rangle$, thus to make a nonadiabatic transition.  
In terms of the time evolution operator
$U_{\infty}=\mathcal T\exp[-i\int_{-\infty}^{\infty}H_{LZ}^{iso}(\tau)d\tau]$,
where $\mathcal T$ is the time ordering operator,
these quantities are defined as 
\bea
P_{s}&\equiv& P_{0\rightarrow 1}=|\langle 1|  U_{\infty}| 0 \rangle |^2,
\nonumber\\
P_{na}&\equiv& P_{0\rightarrow 0} \equiv |\langle 0 | U_{\infty}| 0 \rangle |^2.
\label{eq:def}
\eea
These probabilities add to unity in the closed LZ model.
The nonadiabatic transition probability is given by
\bea
P_{na}^{iso}=\exp(-2\pi\Delta ^2/\alpha).
\label{eq:LZP}
\eea
This exact expression can be derived
based on the solution of the Weber equation \cite{Zen},
or by two different methods \cite{Dykhne,Wittig} that rely on a direct contour integration in the complex $t$-plane. %SSS 

%The survival probability satisfies $P_{s}^c=1-P_{na}^c$.

%-------------------------------------------

\subsection{Open-multistate Landau-Zener model}
\label{model2}

We extend the two-state Hamiltonian (\ref{eq:HLZ}) and describe the ``open-multistate LZ model", 
including a band of parallel-stationary levels,
to account for population relaxation and competing pathways. 
The multistate model includes the original diabats, which are
considered as the system of primary interest (e.g., principal reactant and product, 
the two states of a qubit, or the states of two separated quantum dots), and a discretized 
continuum with $N-1$ levels numbered from $2,3,...,N+1$.
The corresponding Hamiltonian is given by 
\bea
H_{LZ}^{open}(t)&=& H_{LZ}^{iso}(t)+\sum\limits_{j=2}^{N+1}{\epsilon_j\ket{j}\bra{j}}
+\sum\limits_{j=2}^{n+1} {v_{0,j}\ket0\bra{j}}
\nonumber\\
 &+&\sum\limits_{j=2}^{N+1} {v_{1,j}\ket1\bra{j}} + h.c.
%+ \sum\limits_{i=2}^{n+1} {{v_1}\ket{i}\bra1}\nonumber
\label{eq:Hopen}
\eea
The continuum energies extend from $-D$ to $D$, and each of the continuum states $|j\rangle$ couples to both 
diabatic levels $n=0,1$
through the tunneling element $v_{n,j}$. We assume a constant density of states $\rho=(N-1)/2D$, 
and define the hybridization energy as %of the diabats to the continuum as 
\bea
\Gamma_{n}(\epsilon)=2\pi \sum_j |v_{n,j}|^2\delta(\epsilon-\epsilon_j); \,\,\,\,\,\, n=0,1.
\label{eq:Ga}
\eea
%

%----Continuum vs Multistate Interpretations
Since we are considering a discretized continuum, the model (\ref{eq:Hopen}) can be viewed
from different perspectives.
In the first picture the $N-1$ additional states are introduced to capture a true continuum to which the 
diabats are coupled. 
% DDD
While we focus here on a {\it shared} continuum, facilitating indirect transfer between the levels,
in the related  ''lossy LZ model"  the population of the diabats relax to separate reservoirs.
This situations has been
investigated analytically in several works by introducing 
 an imaginary term to the diagonal elements of the Hamiltonian,
see for example Refs. \cite{Akulin,Moyer,Band}. % DDD
This model then becomes complementary 
to the dissipative LZ case where energy relaxation is included \cite{KayanumaPRB,HanggiT0,Thorwart}. 
The second interpretation of the model is simply as a multistate LZ model, 
when the states are allowed to become 
sufficiently dense \cite{DemOsh,Brundobler,Nogo,Sym,Sinitsyn,BT,NBT,GBT1,GBT2}.
Indeed several of the multistate LZ setups may be obtained as special cases of 
the present model:
%SSS
If we set $\Delta = 0$, then we obtain a simplified version of the bow-tie model \cite{BT,NBT} (generalized bow-tie model \cite{GBT1,GBT2}) when we further choose $N=1$ ($N=2$).
% Both the bow-tie model \cite{BT,NBT} and the generalized bow-tie model \cite{GBT1,GBT2} can be seen as 
%special cases to Eq. (\ref{eq:Hopen}) when we allow $\Delta$ to go to zero and set $N=2$ or $N=1$ 
%respectively, albeit when the number of non-stationary states is fixed at 2. 
Alternately, if we nullify both 
$\Delta$ and $\Gamma_1$ (equivalent to removing $\ket1$ from the model entirely) we 
arrive at the Demkov-Osherov model \cite{DemOsh}. 
From a different direction, in the absence of driving ($\alpha=0$) the model can describe population 
transfer  between
two distant quantum wells that are  separated by an intermediate level \cite{CTAP,Kohler}
or by a common reservoir \cite{Gurvitz}. 
In the case of a central reservoir it has been shown that
while the system possesses  a continuum spectrum, it includes bound states in the continuum
responsible for quantum effects such as the formation of 
an entangled state in the spatially separated wells  \cite{Gurvitz}.

The nonadiabatic transition probability is defined here in a similar fashion 
to the closed case (\ref{eq:def}).
%$P_{na}=|\langle 0 \,\, (t\rightarrow \infty)| 0\,\, (t\rightarrow -\infty)\rangle |^2$. 
We evaluate it numerically by propagating the initial state 
$|\Psi(t=-\infty)\rangle=|0\rangle$
through a sequence of short time-evolution operators 
\bea
|\Psi(t+\delta t)\rangle=e^{-iH_{LZ}^{open}(t)\delta t}|\Psi(t)\rangle.
\label{eq:time}
\eea
Numerical parameters are determined following several considerations.
First, we recall that transitions in the closed system are
characterized by the time scale 
$\tau_{tr} \sim \frac{2\pi}{\Delta}$. 
In the open-system model two additional relevant timescales are identified 
as $\frac{2\pi}{\Gamma_0}$ and $\frac{2\pi}{\Gamma_1}$. 
As such, the simulation timestep $\delta t$ must 
satisfy 
\beq
1/\delta t \gg
max\{\Delta,\Gamma_0,\Gamma_1\}.
\label{eq:tstp}
\eeq
Another consideration involves the
initial $t_-$ and final $t_+$ simulation times. These times should be taken long enough so that 
the initial state is prepared (and the final state is reached) far away from the crossing point,
%The LZ model considers a system starting at $t \to -\infty$ and ending at $t \to +\infty$.
%In practice, to reach the asymptotic limit the actual initial and final time, $\pm T$, should satisfy
%
\beq
|t_{\pm}| \gg
 max\left \{\frac{\Delta}{\alpha},\frac{\Gamma_0+D}{\alpha}, \frac{\Gamma_1+D}{\alpha}\right\}.
\label{eq:T}
\eeq

Our main objective here is to understand
the dynamics under the Hamiltonian (\ref{eq:Hopen}) by 
identifying signatures of population relaxation and indirect transfer in the LZ transition rate. 
To achieve this goal
we now present other related-complementary models, which
allow for different effects, energy dissipation and thermal excitation.
%Comparing the dynamics emerging from these two classes of systems will allow us to 
%varaints of the LZ system, when the qubit is coupled to a dissipative environemnt, as we detail next.
%=====================================

\begin{figure}[t]
\includegraphics[scale=0.48]{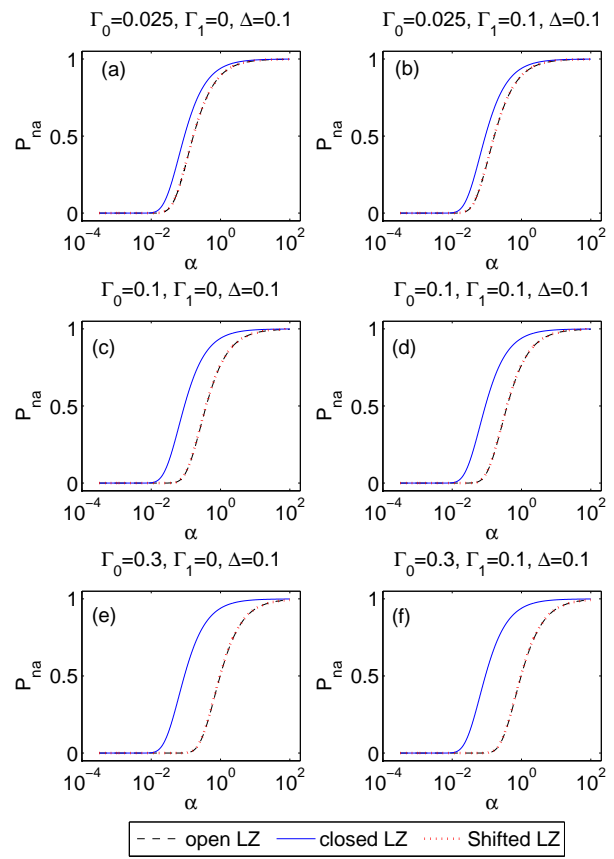}
\caption{
Nonadiabatic transition probabilities in the open-multistate LZ model.
Presented are the
LZ formula from Eq. (\ref{eq:LZP}) (full line),
the probability obtained from simulations with a continuum
of 41 states and $D=1/2$ (dashed),
and $P_{na}^{shift}$ from Eq. (\ref{eq:Shift}) (dotted).
%In all cases $\Delta =0.1$.
The open-multistate model is studied for three coupling strength values
of $\ket0$ and $\ket1$ to the continuum,
denoted by $\Gamma_0$ and $\Gamma_1$ respectively.
}
\label{open1}
\end{figure}

\begin{figure}[t]
\includegraphics[scale=0.45]{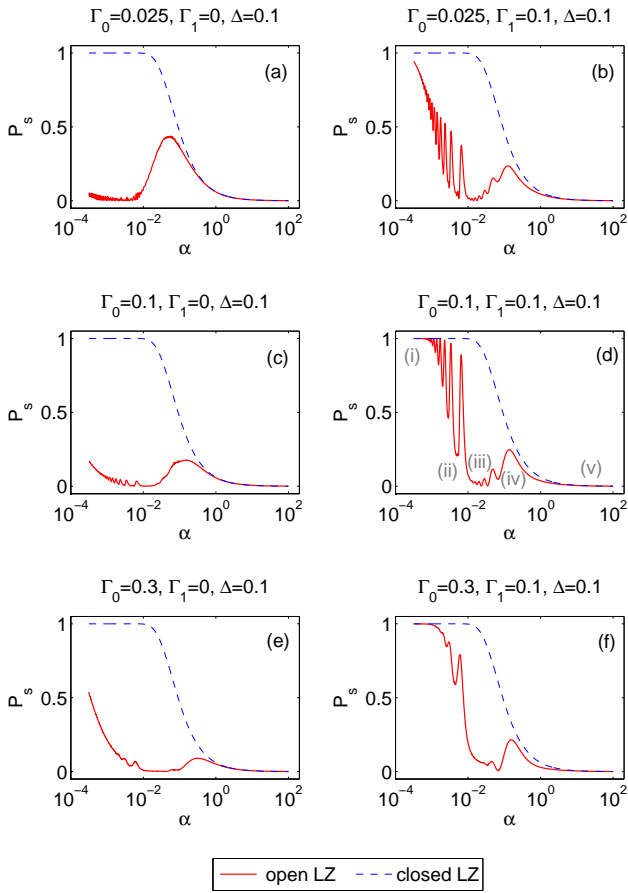}
\caption{Survival probability  in the open-multistate LZ model (full)
compared to the isolated case (dashed).
Parameters are the same as in Fig. \ref{open1}.
In panel (d) we mark the five regions described in the text.}
\label{open2}
\end{figure}

%==================================

%-----------------------------------------------------------------------
\subsection{Stochastic and Dissipative LZ models}
\label{model3}

Nonadiabatic transitions under diagonal energy fluctuations are described with the stochastic model 
\cite{Kayanuma-stoA}, 
\bea
H_{LZ}^{sto}(t)= H_{LZ}^{iso}(t)+ \xi(t)\left(|0\rangle \langle 0| -|1\rangle \langle 1|\right).
\label{eq:HSto}
\eea
The fluctuation energy $\xi(t)$ is 
a Gaussian stochastic random process with a vanishing mean value $\langle 
\xi(t)\rangle=0$. For simplicity, we assume a colored noise with the Ornstein-Uhlenbeck correlation function \cite{OU}
%Markovian correlation function 
%$\langle \xi(t)\xi(t')\rangle =  \kappa^2\delta (t-t')$.
%The standard deviation is given by $\kappa/\sqrt{\delta t}$.
%This could correspond  to
%a stochastic model with a colored
%Ornstein-Uhlenbeck noise, characterized by a finite memory time $\tau_c=\omega_c^{-1}$,
%when the memory time is very short,
%
\bea
\langle \xi(t)\xi(t')\rangle = \frac{\kappa^2}{2\tau_c}e^{-|t-t'|/\tau_c}
\xrightarrow[]
{\tau_c\rightarrow 0} 
\kappa^2 \delta(t-t').
\label{eq:OU}
\eea
Here $\tau_c$ characterizes the memory time of the noise.
When this time is short,  $\kappa^2\tau_c\ll1$ and $\tau_c\ll \tau_{tr}$,
with $\tau_{tr}$ as the characteristic LZ transition time, the correlation function 
can be approximated by the white-noise form.
We simulate the dynamics under (\ref{eq:HSto}) 
by generating the process $\xi(t)$, evolving the dynamics as in Eq. 
(\ref{eq:time}), and averaging the transition probabilities over a large ensemble
for the noise.
% DDD 
Alternatively, one could obtain the transition probabilities by using the time-dependent 
Schr\"odinger-Langevin equation, solving
the two coupled stochastic differential equations \cite{Band}. % DDD

The non-interacting model (\ref{eq:HSto}) has been devised to account for environmental fluctuations in
the LZ behavior. It greatly simplifies the complex
physical situation of level crossing in condensed phases, on surfaces or in solution.
Meanwhile, to include genuine many-body effects we go back to the original model and extend it
in the form of a spin-boson type model,
\bea
H_{LZ}^{diss}(t)&=& H_{LZ}^{iso}(t)  + \sum_{q}\omega_qb_q^{\dagger}b_q
\nonumber\\
&+& \sum_q \lambda_q \left(b_q^{\dagger}+b_q\right)
\left(|0\rangle \langle 0| -|1\rangle  \langle 1| \right).
\label{eq:Hdiss}
\eea
The bath is composed of a collection of harmonic oscillators,
%whose  displacements couple to the polarization.
for which $b_{q}^{\dagger}$ and $b_q$ are bosonic creation and annihilation operators of the $q$ harmonic mode of 
frequency $\omega_q$. 
$\lambda_q$ stands for the interaction energy between the $q$ mode and the diabats' polarization. 
The bath is prepared at the temperature $T$, and its influence on the system is 
characterized by the spectral function  
$J(\omega)=4\pi\sum_{q}\lambda_q^2\delta(\omega-\omega_q)$.
For simplicity, we choose an Ohmic spectral function
\bea
J(\omega)=2\pi K_d\omega e^{-\omega/\omega_c},
\eea
with a cutoff frequency $\omega_c$. The dimensionless Kondo 
parameter $K_d$ quantifies the damping strength. 
The nonadiabatic transition probability and the survival probability, 
from the ground state $|0\rangle$ at $t=-\infty$, are obtained
by evaluating bath-traced density matrix elements
\bea P_{na}&=&tr_B\left[ \langle0| U_{\infty}|0\rangle\langle0| U^{\dagger}_{\infty} 
|0\rangle  \right],
\nonumber\\
P_s&=& 1-P_{na}=tr_B\left[ \langle1| U_{\infty}|0\rangle\langle0| U^{\dagger}_{\infty}       
|1\rangle  \right],
\label{eq:Pnadiss}
%tr_B\left[ \langle0| U_{\infty}\rho(t=-\infty) U^{\dagger}_{\infty} |0\rangle  \right]
\eea 
with the time evolution operator $U_{\infty}=\mathcal T\exp[-i\int_{-\infty}^{\infty}H_{LZ}^{diss}(\tau)d\tau]$.
%The trace is performed over the bath degrees of freedom.
%This quantity can be obtained numerically, using the numerically exact QUAPI technique as explained in \cite{Thorwart}.
The dynamics of this model at nonzero temperature have been the focus of comprehensive studies: 
It has been explored perturbatively-analytically in Ref. \cite{Gefen,Ao,KayanumaPRB,Sun,Ziman}, 
and more recently in Ref. \cite{Thorwart} using a numerically exact technique. 
We do not repeat these investigations here. Rather,
we introduce the Hamiltonian (\ref{eq:Hdiss}) in order to validate
the stochastic model (\ref{eq:HSto}). 
Below we demonstrate that the stochastic description provides results in a qualitative agreement
with the genuine many-body model at weak coupling
for a range of temperatures $T/\Delta=1-50$, see Fig. \ref{QUAPI}.
%Indeed we find that in a certain regime of parameters the transition 
%probability, obtained from the stochastic model, agrees with results based on physical modeling of 
%system-bath interactions. 

We compare dynamics under the models (\ref{eq:HSto}) and (\ref{eq:Hdiss}) 
by noting that  the reorganization energy 
$E_r=\sum_q  4\lambda_q^2/\omega_q=2\omega_c K_d$ 
can be related to the 
variance of the energy fluctuations (in the stochastic model) as
%We now identify the noise memory time by $\tau_c=\omega_c^{-1}$, 
%
$\kappa^2/2\tau_c \propto  E_r T$ \cite{Tsukada,Kayanuma-stoA}.
If we now identify the memory time of the random noise by $\tau_c=\omega_c^{-1}$,
we reach the simple-approximate relation $\kappa\propto \sqrt{K_d T}$,  connecting
the models (\ref{eq:HSto}) and  (\ref{eq:Hdiss}).
The prefactor in this relation should be $\BigO{1}$. In Fig. \ref{QUAPI}
we show that the choice $\kappa=2\sqrt{K_d T}$  consistently provides good agreement between the models.

The probability (\ref{eq:Pnadiss}) is evaluated by time evolving the two-state reduced density matrix using 
QUAPI \cite{Makri}, a numerically exact approach developed for stationary models.
The technique can be naturally extended to simulate time-dependent Hamiltonians; see for example Refs. 
\cite{MakriT,Thorwart}.

%=====================================
\section{Results}
\label{RESULT}

%======================
\begin{figure}[htbp]
\includegraphics[scale=0.5]{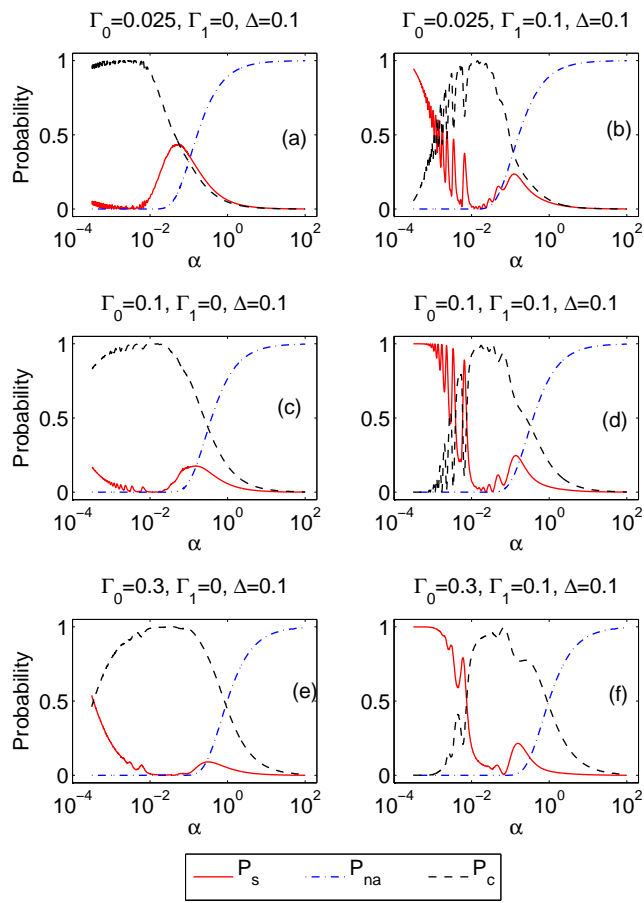}
\caption{Open-multistate model:
Long-time probabilities to occupy the diabats $|0\rangle$ (dashed-dotted) and
$|1\rangle$ (full), or the continuum of states (dashed).
%Parameters are the same as in Fig. \ref{open1}.
}
\label{open3}
\end{figure}

\begin{figure}[htbp]
\includegraphics[scale=0.5]{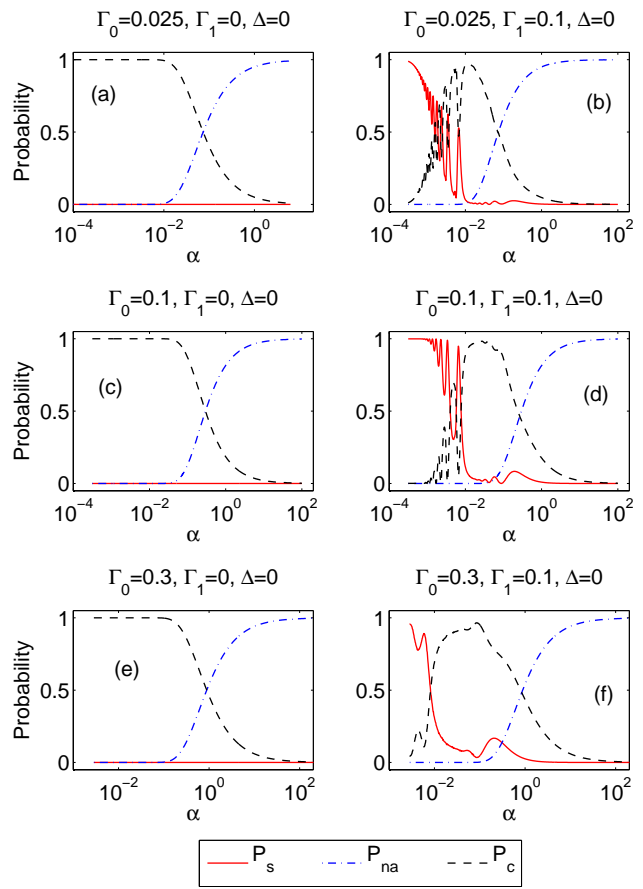}
\caption{Open-multistate model:
Continuum assisted transfer between the diabats with $\Delta=0$.
%Other parameters are the same as in Fig. \ref{open1}.
}
\label{open4}
\end{figure}
%========================

\subsection{Open (multistate) LZ model}
\label{RES1}

We time-evolve the dynamics under the Hamiltonian (\ref{eq:Hopen}) from the initial state 
$|0\rangle$ and explore the long-time 
survival probability in the ground diabatic state, and the probability of
 nonadiabatic transition, as a function of the sweep rate%SSS
 .  
Unless otherwise mentioned we use $\Delta=0.1$. The continuum includes
$41$ states extending $D=\pm 0.5$ with a constant density of states. In our simulations
we set $\Gamma_n$ as an energy independent parameter, and  from this hybridization energy
we resolved the individual coupling strengths using  the relation
$v_{n,k}=\sqrt{\Gamma_n/2\pi\rho}$ from Eq. (\ref{eq:Ga}). 
Similar to the notation typically used in the isolated case, here
$P_s=P_{0\rightarrow1}$ %SSS 
stands for the long-time survival probability in the lowest energy diabat and
$P_{na}=P_{0\rightarrow 0}$ %SSS 
describes the nonadiabatic transition probability.
Furthermore, we define the component residing inside the continuum as $P_c=1-P_s-P_{na}$.

The behavior of $P_{na}$ is displayed in Fig. \ref{open1}. 
It is notable that the value of $\Gamma_1$ does not affect this probability.
Furthermore, the behavior of $P_{na}$ as a function of sweep rate retains the typical LZ form
apart from a shift dependent on $\Gamma_0$.
%for different values of $\Gamma_0$, demonstrating only a shift as it is increased: 
In the case of a dense 
continuum we confirm numerically that this shift corresponds 
to multiplication by an exponential prefactor function,
\bea
P_{na}^{shift}&=&\exp\left(-\frac{4D\Gamma_0}{\alpha} \right)P_{na}^{iso}
\nonumber\\
&=&
\exp\left(-\frac{2\pi\Delta^2+4D\Gamma_0}{\alpha}\right)
\label{eq:Shift}
\eea
We can qualitatively justify this form by considering individual-independent pairwise LZ transitions, 
between the diabat $|0\rangle$ and $|1 \rangle$, and between
$|0\rangle$ and each continuum state. These transitions happen in succession as the level $|0\rangle$
moves closer in energy to other levels. 

More quantitatively, Eq. (\ref{eq:Shift}) can  be justified using
simple semiclassical arguments as follows: Population relaxation into the continuum begins
at time $t_i$ when the adiabatic states reach the lower threshold of the continuum
$-D\sim -\sqrt{\Delta^2+[(\alpha t_i)/2]^2}$. This relaxation is completed at $t_f$,
defined from
 $D\sim \sqrt{\Delta^2+[(\alpha t_f)/2]^2}$. Under the assumption that $\Delta <D$,  
valid in our simulations,
the overall time available for relaxation to the continuum is given by
$\tau_{R}\equiv t_f-t_i\sim 4D/\alpha$.
During this time, population relaxation takes place at the rate $\Gamma_0$.
%SSS
Hence, taking into account the loss to the continuum alone,
the probability to 
remain in the $|0\rangle$ state is given by
the semiclassical expression $\sim e^{-4D \Gamma_0/\alpha}$.
%Following the same reasoning, of independent pairwise transition, at the diabats crossing point
This probability is further modified
by the (independent) LZ pairwise transition when $|0\rangle$ and $|1\rangle$ cross; thus
overall, the total nonadiabatic transition probability follows the multiplicative form
Eq. (\ref{eq:Shift}). 
While this behavior has been observed before in several multistate models 
\cite{DemOsh,Brundobler,Nogo,Sym,Sinitsyn,BT,NBT,GBT1,GBT2,SpinBath},
it is interesting to emphasize %SSS %note
that it is valid not only for well separated states.

%=====================================

\begin{figure}[htbp]
\includegraphics[scale=0.45]{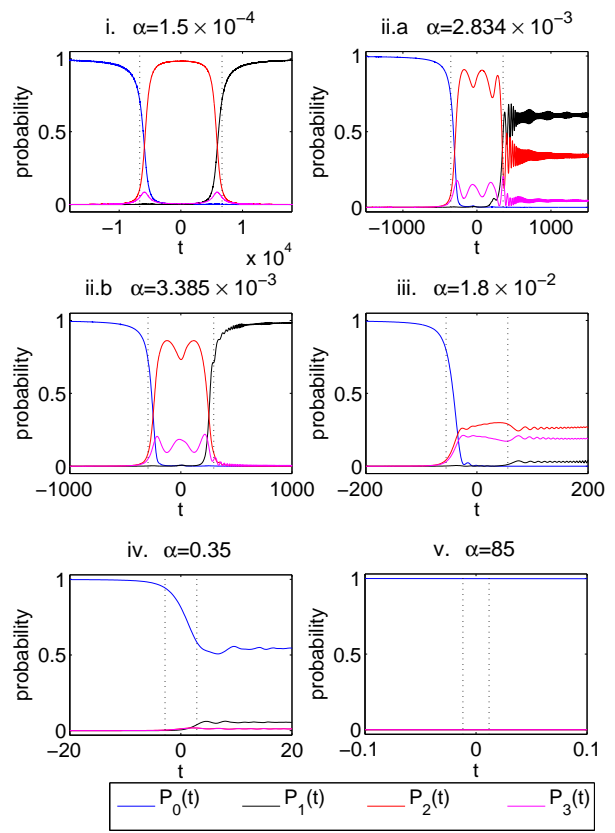}
\caption{The dynamics of the open-multistate Landau-Zener model in the distinct regimes
identified in figure \ref{open2}(d):
(i) adiabatic,
(ii) oscillatory, where we follow the dynamics with a velocity corresponding to
(ii.a) a minimum in $P_s$ or (ii.b) a maximum,
(iii) strong depletion,
(iv) transitionary, and (v) diabatic regime.
Simulations were performed with $\Delta=0$ and $\Gamma_1=\Gamma_0=0.1$.
Plotted are the population of the two diabats, $P_{0}(t)$ and $P_1(t)$
and the population of the two lowest-lying continuum levels $P_2(t)$ and $P_3(t)$.
The dotted lines identify the times $t_{i,f}=\mp 2D/\alpha$, when the adiabatic states
(approximately) touch the lower and upper continuum threshold. %XXX DDD
 }
\label{open5}
\end{figure}
%------------------------------

%\begin{table}[ht]
%\begin{tabular}{|c ||  c| c| c| c| c| c|}
%\hline 
%$\alpha$ & $1.5 \times10^{-4}$ &  $2.9 \times10^{-3}$ & $3.4 \times10^{-3}$ & $1.8 \times10^{-2}$ & $0.35$ & 85 \\
%$t_{i,f}$  & $\pm6.7 \times10^{3}$ & $\pm$350& $\pm$300 & $\pm$55 & $\pm$3 & $\pm$0.01 \\
%\hline
%\end{tabular}
%\caption{Approximate initial and final overlap times with the continuum, $t_{i,f}=\pm 2D/\alpha$,
%useful for interpreting the dynamics displayed Fig. \ref{open5}.}
%\label{tab:alpha}
%\end{table}

%==============

%Survival Probability Results
While nonadiabatic transition probabilities follow the simple shifted LZ form Eq. (\ref{eq:Shift}),
the ground state survival probability
shows a complex behavior due to the influence of the continuum of states as shown in Fig. \ref{open2}. 
Our method reproduces the expected Landau-Zener survival probabilities Eq. (\ref{eq:LZP})
in the closed case (dashed curve). When $\Gamma_n\neq0$, the numerical solution takes on
an entirely different functional form, displaying an oscillatory regime 
as a function of $\alpha$ with alternating maximum and minimum probabilities. 
%Interestingly, depending on the magnitude of $\Gamma_0$ and $\Gamma_1$,the presence of 
%continuum either fascilitates or inhibits adiabatic non-transitions, either extending or shortening the 
%adiabatic regime. 

In the isolated LZ model
the solution (\ref{eq:LZP}) can be divided into three regimes: the adiabatic regime
 $\alpha\ll\Delta^2$, the counter, diabatic regime $\alpha\gg\Delta^2$, and a transitionary
 regime for $1<\alpha/\Delta^2<10^2$. The open LZ model shows two additional 
characteristics at low-intermediate adiabatic rates: an ``oscillatory regime" ($\alpha/\Delta^2\sim 0.1$),
 followed by a ``strong depletion" phase $\alpha/\Delta^2\sim 1$.
We assign these five distinct regimes as shown in Fig. \ref{open2}(d) as
(i) adiabatic, (ii) oscillatory, (iii) strong depletion, (iv) transitionary,  and
(v) diabatic.

%Full Leaking Results
A more complete picture of the open model may be reached by simultaneously considering the probabilities 
of the system being found in $\ket0$, $\ket1$, or in the continuum states 
$\left\{\ket2,\ldots,\ket{N+1}\right\}$. In Fig. \ref{open3} we plot the long-time probabilities 
$P_s=P_{0\rightarrow 1}$, %SSS
$P_{na}=P_{0\rightarrow 0}$, %SSS
and $P_c=\sum_{j=2}^{N+1}P_{0\rightarrow j}$, for residing in the continuum. 
We make the following observations:
(A) The oscillatory behavior of 
$P_s$ is strong when $\Gamma_1\neq0$;
increasing the value of $\Gamma_0$ largely affects these features.
(B) Both the oscillatory regime and the strong depletion regime result from %some velocities favoring
the system remaining in the continuum rather than making a further %SSS
nonadiabatic transition to
the other diabat for some $\alpha$ values.
Furthermore, we show in Fig. \ref{open4} that the oscillatory structure is in fact preserved
 for $\Delta=0$, even maintaining nearly %SSS
 the same values for the transition velocities. 
(C) When $\Gamma_1=0$, the survival probability at low values of $\alpha$ increases with $\Gamma_0$,
which is an evidence for the presence of cotunelling processes in the system, see panels (a),
(c) and (e) in Fig. \ref{open3}.
(D) The shift in the nonadiabatic transition probabilities as described beneath Eq. (\ref{eq:Shift})
results from an increased occupation in the continuum, rather than from an increase in probability 
of finding the system in the other diabatic state.

%Open Reults
%In addition to the leaking and dissipative cases, a single parameter set was studied for a fully open system 
%with both the presence of the equilibrium and the application of Gaussian noise to the diabatic potentials. 
%This system showed one particularly interesting characteristic in comparison to the dissipative system 
%without a continuum. Namely, the non-adiabatic transition probability in the open system seems to be 
%entirely shielded from the effects of the Gaussian noise and only displays the shift characterized in the 
%analysis of the purely leaking system. This shielding effect can be seen in figure \ref{fig:LZlkSto}. 

%We explore nonadiabatic transition probabilities in the LZ model variants, particularly separating
%signatures of population transfer (open LZ model) from energy dissipation at infinite temperature
%(stochastic LZ model) and finite temperature (dissipative LZ model). In the calculations presented below the
%initial state is the ground state, $|0\rangle$.

%==============

To understand the complex behavior of the survival probability in the different regimes
we 
%closely consider the dynamics of the system about the transition time at different values of $\alpha$.
%%
%SSS
choose several specific values of $\alpha$ from Fig. \ref{open4} and examine the time evolution
across the transition region in detail.
This is illustrated in figure \ref{open5} using parameter values
 $\Delta=0$ and $\Gamma_0=\Gamma_1=0.1$.
Recall 
%that  we have found in Fig. \ref{open4} 
as discussed above
that the oscillatory regime occurs here
independently of $\Delta$.
Inspecting Fig. \ref{open5} we confirm that the continuum begins to participate in the
dynamics,
in agreement with the semiclassical estimate, 
when $\epsilon_0$ enters the continuum at $t_i=-2D/\alpha$. 
At $t_f=2D/\alpha$ the diabats depart from the continuum, and their populations
begin to relax to the respective asymptotic values. 
These times are marked by dotted lines in Fig. \ref{open5}.

Inspecting  
the behavior at very low sweep velocities 
(Region i in Fig. \ref{open5}) %the oscillations of populations in the continuum states)
% have sufficient time to damp out to the asymptotic
%limit, in addition to further favoring the 
we find that the population is transferred  
almost entirely from $\ket0$ to the lowest-energy continuum
state $\ket2$ at the beginning of the transition region.
 Furthermore, the oscillations in the dynamics, after the transition,
have sufficient time to damp out. At the last stage, around the time $t_f$ the continuum level 
$|2\rangle$ slowly crosses the diabat
$|1\rangle$ and its population is transferred entirely to the state $\ket1$. 
This creates a new, $\Delta$-independent adiabatic regime, governed by 
a continuum-assisted adiabatic transition, explaining the observation of a finite adiabatic survival probability even with $\Delta = 0$.

We now aim to explain the oscillatory Region ii. 
%Upon considering the behavior of individual continuum states we observe
%that all states display an oscillation in time at one specific frequency, independent of $\alpha$. 
%
%SSS
In Fig. \ref{open5}(ii.a) and (ii.b) we observe that the time evolution of the continuum states is itself oscillatory, 
with a specific frequency that is $\alpha$-independent.
Maxima in the asymptotic survival probability occur when the continuum-interaction duration
$\tau_R=4D/\alpha$
%, during which the energy of the adiabats are embedded within the continuum,
is commensurate with the period of this frequency.

Note that the oscillations observed in Fig. \ref{open5} are not unique to the continuum states,
 rather they are a characteristic of the LZ transition,
and are seen near the transition region of any pairwise crossing before damping out to the asymptotic
limit. In the closed LZ model these oscillations decrease both in period and amplitude as they are
damped to a constant value. In addition, their frequency is dependent on the coupling constant $\Delta$.
The decrease in period is not observed in the oscillatory regime in the present case as $\tau_R$ is
sufficiently short, such that the period damping is insignificant. 
Note that at these  low sweep velocities
significant population is transferred to the lowest energy states of the continuum, 
resulting in a very low transfer of population
to higher continuum states in subsequent crossings.
This explains the similar behavior of the model either with or without  $\Delta$, 
as demonstrated in the comparison between Fig. \ref{open3} and Fig. \ref{open4}.

Next, we explain the depletion region (iii) when $\alpha/\Delta ^2\sim 1$. For such velocities
the continuum levels  approach their stationary-asymptotic
limit during  $\tau_R$, as in the usual two-state LZ dynamics.
For even higher velocities $\alpha/\Delta ^2> 1$, 
all continuum states gain population, albeit to a very small extent,
thus the survival probability approaches (region iv) and then
recovers (region v) the standard LZ  curve.

%The shielding effects observed in the open Landau-Zener system can be explained in terms of the order in 
%which the states interact with each other in the sweep velocity range in which the effects of the Gaussian 
%noise are significant. In this range, the $\ket0$ diabat begins to interact with the continuum before it 
%interacts with the $\ket1$ diabat. Thus, the probability transfer between the $\ket0$ and $\ket1$ diabat 
%experiences a decreased effect of the Gaussian noise as the continuum behaves as a buffer for population 
%transition. 

%====================

\subsection{Stochastic and dissipative LZ models}
\label{RES2}

In this section we explore principal 
signatures of energy dissipation and thermal excitations in the LZ transition rate
and show that these characteristics can
be separated from the effect of competing channels as described in Sec. \ref{RES1}.

\begin{figure}[thb]
\includegraphics[scale=0.38, angle=0, origin =c]{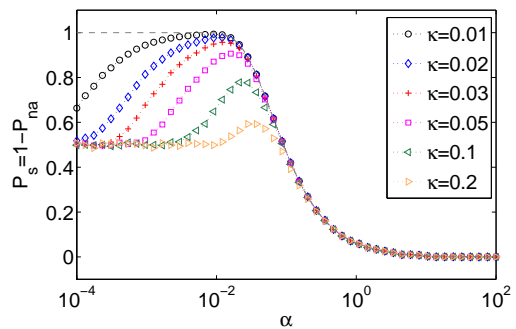}
\caption{Stochastic LZ model:
 Probability for survival $P_{s}$, to stay on the ground state at long time,
$\kappa=0.01,0.02,0.03,0.05,0.1,0.2$
top to bottom. The dashed line stands for the closed model, $\kappa=0$.
%(b) Complementary nonadiabatic probability. 
In all cases $\Delta=0.1$. }% and $\delta t=2\pi/\Delta/100$.}
\label{FigSto}
\end{figure}

We simulate the LZ model with Markovian-Gaussian
energy fluctuations, Eq. (\ref{eq:HSto}), and display these results in Fig. \ref{FigSto}. 
We use $\Delta=0.1$ and apply the noise only when the diabats satisfy the condition
$|\epsilon_0(t)-\epsilon_1(t)|<\omega_c$ with $\omega_c=1$.
The characteristic transition time is longer than the noise decorrelation time,
$\tau_{tr}\sim 2\pi/\Delta \gg \tau_c$, $\tau_c=\omega_c^{-1}\sim 1$. Further, since
we only work in the weak coupling limit,
$\kappa^2\tau_c\ll 1$,
we can perform our simulations using a Markovian (delta function) correlation function.
The behavior observed in Fig. \ref{FigSto} is in agreement with Kayanuma's results
\cite{Kayanuma-stoA}, displaying a nonmonotonic behavior with an optimal velocity for 
survival in the low $\kappa$ limit, while reducing to the LZ formula (scaled by a factor 
of $\frac{1}{2}$) when $\kappa$ is made larger. Note that $P_{na}=1-P_s$ for the stochastic model, since population leakage is not allowed.

\begin{figure}[thb]
\includegraphics[scale=0.38, angle=0, origin =c]{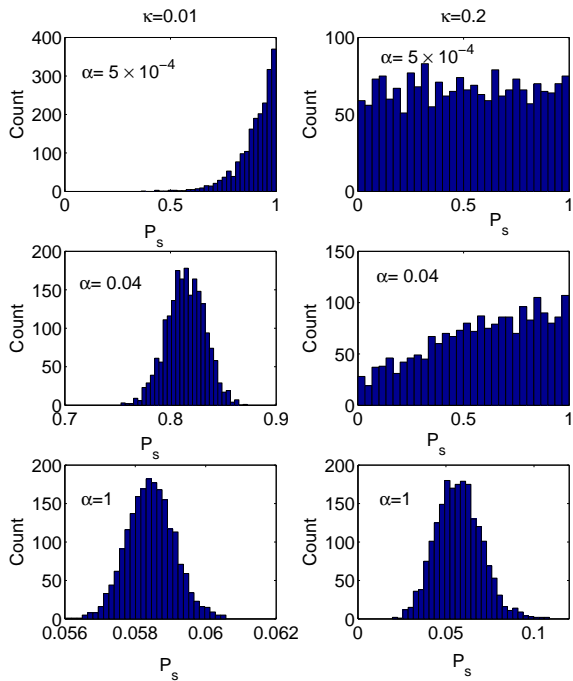}
\caption{Stochastic LZ model:
Histogram of the survival $P_{s}$ probability, to stay on the ground state at long time,
$\kappa=0.01$ (left panels) and $\kappa=0.2$ (right panels),
with $\alpha=5 \times 10^{-4}$, 0.04 and 1, top to bottom.
In all cases $\Delta=0.1$. }
\label{Figprob}
\end{figure}

%DDD
Fig. \ref{Figprob} shows histograms of $P_s$ at different sweep rates, and for small
($\kappa=0.01$) and large ($\kappa=0.2$) noise values, based on 2000 stochastic realizations.
As expected, in the diabatic (high velocity) regime weak and strong noise processes lead to similar results
for the probability distribution of the survival probability.
However, at intermediate-to-small sweep rates ($\alpha/\Delta^2 <1$) we find marked deviations:
For large $\kappa$ the distribution seems uniform in the range $[0,1]$, trivially providing the mean $P_s(\alpha=10^{-4})\sim 0.5$.
In contrast, at
small $\kappa$ the distribution seems to follow an exponential form.
%Analogous results were obtained when implementing dephasing processes within the
%Schr\"odinger–Langevin equation approach \cite{Band}.
%%% DDD

%===============================================================

% DDD

\begin{figure}[htb]
\includegraphics[scale=0.4, angle=0, origin =c]{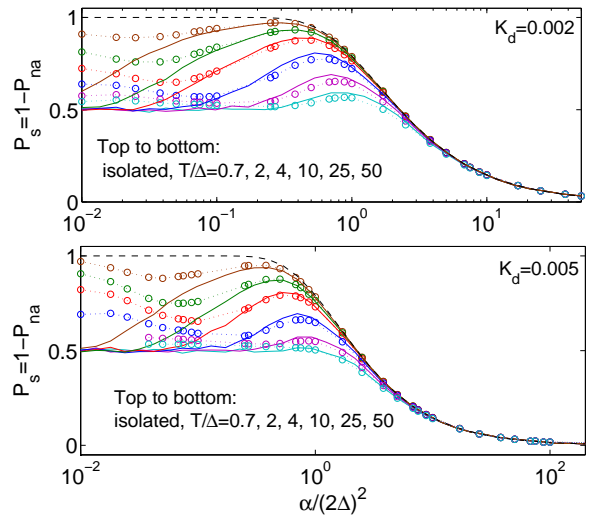}
\caption{LZ survival probability in the stochastic model (full) and the dissipative case
for $T/\Delta=0.7,2,4,10,25,50$ (top-to-bottom, $\circ$).
The dashed line stands for the LZ formula (\ref{eq:LZP}).
%(b) Zoom over the region $\alpha/\Delta^2>0.1$
%where (nontrivial) agreement between the two simulations is observed.
(a) Kondo parameter $K_d=0.002$, (b) $K_d=0.005$.
Other QUAPI parameters are %$K_d=E_r/2\omega_c=0.0016$, 
$\omega_c=20\Delta$ and $\Delta=0.1$.
In the stochastic description the variance was prepared from the relation
$\kappa= 2\sqrt{K_d T}$.
}
\label{QUAPI}
\end{figure}

%====================

The stochastic model emulates the effect of an environment within a fluctuating field,
and one should question whether this type of modeling is (at least) qualitatively correct.
This issue is addressed in Fig. \ref{QUAPI}, in which
%We  show that for temperatures $T/\Delta\sim 1-50$
%stochastic simulations provide results in a good qualitative agreement with 
%the fundamental model,
%(\ref{eq:Hdiss}) in which a many-body environment in included
%for $\alpha/\Delta^2>0.1$.
%
the LZ dynamics of the dissipative model are studied numerically-exactly using the QUAPI technique
\cite{Makri}, as explained in Ref. \cite{Thorwart}.
We used  a large cutoff $\omega_c= 20\Delta$, a small Kondo parameter $K_d=0.002-0.005$
and a range of intermediate-high temperatures $T/ \Delta=1-50$.
QUAPI simulations are aligned with the stochastic model (\ref{eq:HSto})
by employing the relation $\kappa =2\sqrt{K_d T}$, see Sec. \ref{model3}.
This translates to noise amplitudes extending the range $\kappa=0.07-1$.
%with $\gamma$ as a numerical value of $O(1)$. In practice, we found that the value $\gamma=2$
%The numerical prefactor was chosenproduces results in good agreement with the dissipative model, 
%for a broad range of temperature,
%$T/\Delta=0.7-50$ for different values of $K_d$.
%
We display our results in Fig. \ref{QUAPI} in the same format as in  Ref.
\cite{Thorwart}, to allow for a quick comparison. 
We find that the peak in the survival probability with velocity is correctly captured (position, height) 
within the stochastic model for high-to-intermediate temperatures in the $T/ \Delta=1-50$ range.
At small sweep velocities $\alpha/\Delta^2< 0.1$  marked deviations appear:
Stochastic simulations approach the probability $P_s=1/2$, representing an infinite-temperature bath,
while the QUAPI method indicates that
the two levels adjust their population to the thermal distribution as 
dictated by the temperature and the instantaneous energy gap. 
%This simple comparison 
%allows us to conclude that the principal effect of the heat bath
%is to reduce the survival probability from its noninteracting value. 
For higher velocities,  $\alpha/\Delta^2> 0.1$,
%the two states are swept fast enough 
 relaxation with respect to the instantaneous gap is not reached, and
the effect of the temperature and the coupling strength can be apparently captured within a single
parameter $\kappa$, characterizing energy fluctuations.

%===============================================================

%We thus conclude that
%for temperatures $T/\Delta\sim 1-50$ and for nonadiabatic values $\alpha/\Delta^2>0.1$
%stochastic simulations provide results in a good qualitative agreement with 
%the fundamental model (\ref{eq:Hdiss}).
It is interesting to note that the two techniques (QUAPI and stochastic simulations)
converge in counter manners.
QUAPI is easy to converge at high temperatures when the bath decorrelation time is short \cite{Makri}.
In contrast, at high temperatures the noise processes suffers from a high 
variance, $\kappa^2\propto T$, thus stochastic simulations 
necessitate significant averaging.

%===============================================================

\subsection{Multistate-stochastic model}
\label{RES3}

We have learned in Sections \ref{RES1} and \ref{RES2} that
population decay from the diabats to other states and energy dissipation processes
have distinct effects on the LZ tunneling
probabilities. When energy exchange with an environment  is permitted, in the form of a stochastic noise,
$P_{na}$ acquires a finite value at low sweep rates and a minimum value of $\frac{1}{2}$ in the high $\kappa$ limit,
which essentially eliminates the adiabatic limit, see Fig. \ref{FigSto} (with $P_{na}=1-P_s$). 
In contrast, when other channels are included $P_{na}$ simply
shows a positive shift in the velocity coordinate (recall Fig. \ref{open1}).
 More interesting is the effect of the
environment on the survival probability.
Allowing energy dissipation in the form
of stochastic noise
 introduces relatively straightforward non-monotonic behavior
 at low $\kappa$ values, and
simply scales the LZ formula at high $\kappa$,  Fig. \ref{FigSto}. In contrast,
the introduction of a resolved continuum displays
rich features at low-intermediate sweep velocities, as in Fig. \ref{open2}. 
It is essential to probe whether
these fine details would survive when interactions with a dissipative environment are in effect.
%In addition, the phenomena observed in a dissipative
%environment occur as a result of energetic factors, while those that occur in the presence of a continuum
%occur as a result of the dynamics of the transitions.

\begin{figure}[t]
\includegraphics[scale=0.38, angle=0, origin =c]{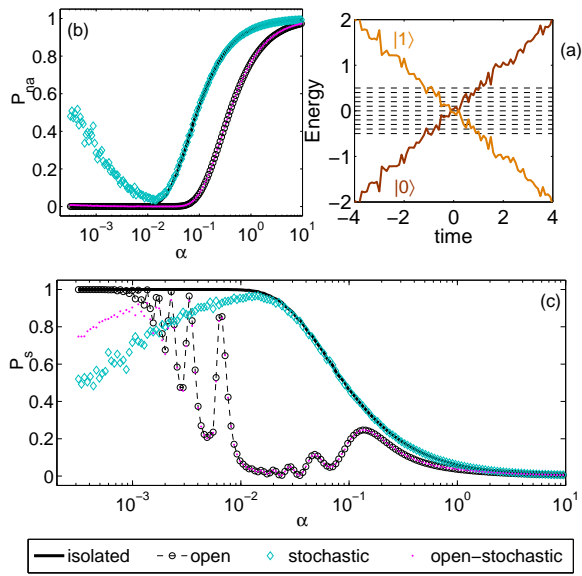}
\caption{(a) Scheme of the open-stochastic LZ model.
(b) Nonadiabatic transition probability $P_{0\rightarrow 0}$ and
(c) Survival probability $P_{0\rightarrow 1}$ for the
 isolated LZ model [Eq. (\ref{eq:HLZ}), full],
open-multistate model [Eq. (\ref{eq:Hopen}), $\circ$],
stochastic LZ model [Eq. (\ref{eq:HSto}), $\diamond$],
and the open-stochastic case (dotted).
Parameters are
 $\Delta=0.1$, $\Gamma_0=\Gamma_1=0.1$, $D=\pm 0.5$ and $\kappa=0.03$.
}
\label{openS}
\end{figure}

%Amro: send me pls the .ps + run more simulations.
We address this issue by adjoining to
the open-multistate model Eq. (\ref{eq:Hopen}) a stochastic noise term affecting the energies of the diabats,
as in Eq. (\ref{eq:HSto}); see the scheme in Fig. \ref{openS}(a).
Results are displayed in Fig. \ref{openS}(b)-(c) for 
$\kappa=0.03$. This value could correspond to $K_d=0.001$ 
and $T=0.25$ in the genuine many-body model. 
For such parameters stochastic simulations are expected to be physically meaningful  
beyond the strict adiabatic limit, once $\alpha/\Delta^2>0.1$, see Fig. \ref{QUAPI}.

We find that the nonadiabatic transition probability in the
open system is entirely shielded from the application of the
 noise, and it only displays the shift characteristic to the effect of the continuum
on the original LZ behavior as in Eq.
(\ref{eq:Shift}). 
The survival probability is susceptible to the noise in the adiabatic regime leading to some loss,
but the oscillations at low sweep rates and other open-system features are all excellently
 protected from the noise in comparison to the case without the continuum.
%====================

%================

%-----------------------------------------------------------------

\section{Conclusions}
\label{CONC}

Nonadiabatic level crossings are affected
by the presence of other channels, discrete or dense, and by energy dissipation and 
thermal excitation processes induced by the surrounding environment.
When inspecting the  ground state survival probability and the nonadiabatic transition probability
and noting deviations from Eq. (\ref{eq:LZP}), can we pinpoint the physical mechanism
responsible for these deviations?

To address this problem we have focused in this paper on the open-multistate LZ model of 
Fig. \ref{Fig1}(a)
including two diabats and a shared discretized continuum.
Using numerical simulations, we have developed a simple analytic expression for the 
probability of nonadiabatic transition 
Eq. (\ref{eq:Shift}), which is supported by
semiclassical considerations of the relaxation time to the continuum.
This expression preserves the functional form of the closed LZ formula while simply applying 
a shift along the sweep velocity  coordinate. 
In contrast, the ground state survival probability at long time
 manifests an entirely different non-monotonic functional form  
in the presence of a continuum, with several features that are not 
present in the original LZ model. The most striking of these features is the presence of an oscillatory 
regime at low sweep velocities and the onset of a novel continuum-facilitated adiabatic regime. % Fig. \ref{open5}(i).
%Finally, 
%we have provided evidence for numerical predictions that the presence of a continuum may act to shield the 
%non-adiabatic transition probability from the effect of Gaussian noise, leaving a definite, easily 
%determined shift. This may provide fertile ground for development of novel shielding techniques for Quantum 
%TLS's. 
%
We have explained these features by referring to the dynamics of the original closed
 LZ problem and by considering 
the individual transitions between the various states as quasi-independent. 
Specifically, a continuum-facilitated 
adiabatic regime replaces the original $\Delta$-dependent regime, and operates through a different 
mechanism: The probability is transferred from the initial diabatic state to the continuum,
 and then back to the other diabatic state as illustrated in Fig. \ref{open5}(i). %SSS 
%This contrasts to the typical adiabatic survival mechanism in the closed LZ 
%model where the probability is directly transferred to the other diabat at their crossing point. 
Meanwhile, the oscillatory regime has been explained with respect to the commensurability of the relaxation time and the period  of the oscillations observed in the continuum occupation probabilities %SSS
near the transition time. 
% and the magnitude of the independent  LZ transition at each independent crossing. 

We have complemented the study of the multistate model by considering two other environment-affected
LZ models:
the stochastic LZ model in which  the energy of the diabats
 fluctuates around the original value, 
and a more fundamental model in which the diabats couple to a harmonic
environment at finite temperature. 
These models allow for energy exchange processes, reflected by a maxima
in the survival probability around $\alpha/\Delta^2\sim1$, and a modification to the 
deep adiabatic limit, to approach a thermal distribution ratio.
Realizations are anticipated in studies of reactions on or close to surfaces \cite{Tully1, Tully2},
in the field of LZ interferometry, a sensitive tool that can decipher the details of the 
environment affecting a double-quantum dot system \cite{KohlerE},
and in other adiabatic devices,
where gated quantum dots may indirectly couple  \cite{CTAP,Kohler,Gurvitz},
 to facilitate population transfer.

In conclusion, in order to identify factors affecting transitions at avoided level crossings one
should review both the nonadiabatic transition probability and the survival probability, 
as they may reveal distinctive features when multichannels are involved.
%Specifically, the nonadiabatic transition probabilty reveal only a 
%simple shift in the velocity coordinate  when competting channels participate.
%However, when energy relaxation is allowed a minima will develop around 
%$\alpha/\Delta^2\sim 1$, and the probability will approach a thermal distribution in the adiabaic limit.
%
%One could clearly
%separate population decay from energy relaxation when studying the survival probability,
%while the nonadiabatic transition probability show less salient features of the environment.
%
%In the presence of competting channels Fig. \ref{open2} reveals oscillatory features
%for $\alpha/\Delta^2\sim 1$ when $\Gamma_1\neq0$, followed by a strong depletion area.
%When only $\Gamma_0$ is nonzero the survival probability is strongly depleted 
%for $\alpha/\Delta^2< 1$, as long as $\Gamma_0/\Delta<1$. 
%Under energy fluctuations, the survival probability is complementary to $P_na$,
%showing a maxima around $\alpha/\Delta^2\sim 1$ and an approach to a thermal equilibrium
%when $\alpha/\Delta^2\ll1$.
%
We have also provided numerical evidence that the presence of a continuum may act to shield the 
nonadiabatic transition probability from the effect of a Gaussian noise, leaving only a definite, 
predictable shift. This may provide a fertile ground for the 
development of novel shielding techniques in quantum adiabatic computing \cite{QAdi,AdComp}.

%=========================
\begin{acknowledgments}
We thank Naomichi Hatano and Takashi Oka for insightful discussions.
A. D. acknowledges support from the NSERC USRA program. 
The work of S. G. was supported by a JSPS Research Fellowship
and by the CQIQC visiting program.
L. S.  was supported by an Early Research Award of D. S.
D. S. acknowledges support from an NSERC discovery grant.
\end{acknowledgments}
%==========================

%-----------------------------


\begin{thebibliography}{92}

%=====================
% old LZ
\bibitem{Zen}
C. Zener, Proc. R. Soc. A. {\bf 137}, 696 (1932).

\bibitem{Land}
L.D. Landau, Phys. Z. Sowjetunion {\bf 2}, 46 (1932).

\bibitem{Stueck}
E.C.G. Stueckelberg, Helv. Phys. Acta {\bf 5}, 369 (1932).

\bibitem{Maj}
E. Majorana, Nuovo Cimento {\bf 9}, 43 (1932).

%====================
% appl

\bibitem{Reaction}
A. Nitzan,
{\it Chemical Dynamics in Condensed Phases}
 (Oxford University Press, Oxford 2006).


\bibitem{ColdRev}
T. K\"ohler, K. Goral,  and P, S. Julienne,
Rev. Mod. Phys. {\bf 78}, 1311 (2006).
%Production of cold molecules via magnetically tunable Feshbach resonances

\bibitem{QED1}
%I. Chiorescu {\it et al.}, 
I. Chiorescu, P. Bertet, K. Semba, Y. Nakamura, C. J. P. M. Harmans, and  J. E. Mooij,
Nature  {\bf 431}, 159 (2004).
% Coherent dynamics of a flux qubit coupled to a harmonic oscillator

\bibitem{QED2}
%A. Walraff {\it et al.}, Natue (London) {\bf 431}, 162 (2004).
%Strong coupling of a single photon to a superconducting qubit using circuit quantum electrodynamics, 
A. Wallraff, D.  I. Schuster, A. Blais, L. Frunzio, R.- S. Huang, J. Majer, S. Kumar, S.  M. Girvin, 
and R.  J. Schoelkopf, Nature {\bf 431}, 162 (2004).

\bibitem{nanmag}
W. Wernsdorfer and R. Sessoli, Science {\bf 284}, 133 (1999).

\bibitem{BEcond}
A. Zenesini, H. Lignier, G. Tayebirad,  J. Radogostowicz, D. Ciampini, 
R. Mannella, S. Wimberger, 
O. Morsch, and E. Arimondo,
%A. Zenesini {\it et al.}, 
Phys. Rev. Lett. {\bf 103}, 090403 (2009).
% Time-Resolved Measurement of Landau-Zener Tunneling in Periodic Potentials

\bibitem{Oka}
T. Oka,
Phys. Rev. B {\bf 86}, 075148 (2012).

\bibitem{bipartite}
S. Longhi and G. Della Valle,
Phys. Rev. A {\bf 86}, 043633 (2012).

\bibitem{QAdi}
E. Farhi, J. Goldstone, S. Gutmann, and M. Sipser,
%“Quantum computation by adiabatic evolution”, 2000. 
arXiv:quant ph/0001106


\bibitem{AdComp}
E. Farhi, J. Goldstone, S. Gutmann, J. Lapan, A.
Lundgren, and D. Preda, Science {\bf 292}, 472 (2001).
%“A quantum adiabatic evolution algorithm applied to random instances of an NP complete problem”.


%----------------

\bibitem{LZS}
S.N. Shevchenko, S. Ashhab, and F. Nori, Phys. Rep. {\bf 492}, 1 (2010).
% LZ spectroscopy

%-------------------

\bibitem{DemOsh}
%Yu.N. Demkov, V.I. Osherov, J. Exp. Theor. Phys. {\bf 24}, 916 (1938).
Y. N. Demkov and V. I. Osherov, Zh. Exsp. Teor. Fiz. {\bf 53}, 1589 (1967)
[Sov. Phys. JETP {\bf 26}, 916 (1968)].


% multistate 
\bibitem{Brundobler}
S. Brundobler and V. Elser, J. Phys. A Math. Gen. {\bf 26}, 1211 (1993).
% S-matrix for generalized Landau-Zener problem 

\bibitem{Nogo}
N. A. Sinitsyn, J. Phys. A Math. Gen. {\bf 37}, 10691 (2004).

\bibitem{Sym}
N. A. Sinitsyn, Phys. Rev. A {\bf 87}, 032701 (2013).
%Landau-Zener transitions in chains 

\bibitem{Sinitsyn}
N. A. Sinitsyn, Phys. Rev. Lett. {\bf 110}, 150603 (2013).
% Nonadiabatic Transitions in Exactly Solvable Quantum Mechanical Multichannel Model:
%Role of Level Curvature and Counterintuitive Behavior

% Bow tie models
\bibitem{BT}
C. Carrol and F. T. Hioe, J. Opt. Soc. Am. B {\bf 2}, 1355 (1985).

\bibitem{NBT}
V. N. Ostrovsky and H. Nakamura,  J. Phys. A: Math. Gen. {\bf 30}, 6939 (1997).
% Exact analytical solution of the N-level Landau - Zener-type bow-tie model 

\bibitem{GBT1}
Yu. N. Demkov and V. N. Ostrovsky, Phys. Rev. A {\bf 61}, 032705 (2000).
% Introduces the generalized bow-tie model and solves it heuristically

\bibitem{GBT2}
Yu. N. Demkov and V. N. Ostrovsky, J. Phys. B At. Mol. Opt. {\bf 34}, 2419 (2001).
% The exact solution of the multistate Landau-Zener type model: the generalized bow-tie model 

\bibitem{SpinBath}
N. A. Sinitsyn and N. Prokof’ev, Phys. Rev. B {\bf 67}, 134403 (2003).
% Nuclear spin bath effects on Landau-Zener transitions in nanomagnets
% coupling between the electronic and nuclear spins

\bibitem{Akulin} %DDD
V. M. Akulin and W. P. Schleich,
Phys. Rev. A {\bf 46}, 4110 (1992).
% Landau-Zener transition to a decaying level


\bibitem{Moyer} %DDD
%Quantum transitions at a level crossing of decaying states
C. A. Moyer,
Phys. Rev. A {\bf 64}, 033406 (2001).

\bibitem{Band} %DDD
%The Landau–Zener Problem with Decay and with Dephasing
Y. Avishai and Y. B. Band,  arXiv:1311.3919.


%-------------------
% genuine bath

\bibitem{Tsukada}
M. Tsukada, J. Phys. Soc. Jpn {\bf 51}, 2927 (1982).
% Theory of Non-Adiabatic Processes of Adsorbates
%  interaction with the surface heat bath.

\bibitem{Gefen}
Y. Gefen, E. Ben-Jacob, and A. O. Caldeira, Phys. Rev. B {\bf 36}, 2770 (1987).

\bibitem{Ao}
P. Ao and J. Rammer,
Phys. Rev. B {\bf 43}, 5397 (1991) 
%Quantum dynamics of a two-state system in a dissipative environment

\bibitem{Stern}
E. Shimshoni and A. Stern, Phys. Rev. B {\bf 47}, 9523 (1993).
% Dephasing of interference in Landau-Zener transitions

\bibitem{KayanumaPRB}
Y. Kayanuma and H. Nakayama, Phys. Rev. B {\bf 57} 13099 (1998).
%Nonadiabatic transition at a level crossing with dissipation
% phonons

\bibitem{Sun}
V. L. Pokrovsky and D. Sun, Phys. Rev. B {\bf 76}, 024310 (2007).

% DDD
\bibitem{spin}
D. A. Garanin, R. Neb, and R. Schilling,
 Phys. Rev. B {\bf 78}, 094405 (2008).
%Effect of environmental spins on Landau-Zener transitions


\bibitem{Ziman}
%Temperature Can Enhance Coherent Oscillations at a Landau-Zener Transition
R. S. Whitney, M. Clusel, and T. Ziman, Phys. Rev. Lett. {\bf 107}, 210402 (2011).


\bibitem{HanggiT0}
M. Wubs, K. Saito, S. Kohler, P. H\"anggi, and Y. Kayanuma, Phys. Rev. Lett. {\bf 97}, 200404 (2006);
K. Saito,  M. Wubs, S. Kohler, Y. Kayanuma, and P. H\"anggi, Phys. Rev. B {\bf 75}, 214308 (2007).
%D. Zueco, P. Hänggi, and S. Kohler, New J. Phys. 10, 115012 (2008).
% zero temperature.
%Dissipative Landau-Zener transitions of a qubit: Bath-specific and universal behavioto

\bibitem{Thorwart}
P. Nalbach, M. Thorwart,  Phys. Rev. Lett. {\bf 103}, 220401 (2009),
Chem. Phys. {\bf 375}, 234 (2010).
% LZ transitions in a dissipative environemnts: Numerically exact results

\bibitem{LeHur}
P. P.  Orth, A. Imambekov, and K. Le Hur, Phys. Rev. A {\bf 82}, 032118 (2010);
Phys. Rev. B {\bf 87}, 014305 (2013).

\bibitem{Tully1}
A. M. Wodtke, J. C. Tully, and D. J. Auerbach,
Int. Reviews in Phys. Chem., {\bf 23}, 513 (2004).
% Electronically non-adiabatic interactions of molecules at metal surfaces: 
%Can we trust the Born–Oppenheimer approximation for surface chemistry?

\bibitem{Tully2}
%Dynamical Steering and Electronic Excitation in NO Scattering from a Gold Surface
N. Shenvi, S. Roy, and J. C. Tully,
Science {\bf 326}, 829 (2009).

\bibitem{CTAP}
A. D. Greentree, J. H. Cole, A. R. Hamilton, and L. C. L. Hollenberg,
Phys. Rev. A {\bf  70}, 235317 (2004).
% Coherent electronic transfer in quantum dot systems using adiabatic passage

\bibitem{Kohler}
%Steady-State Coherent Transfer by Adiabatic Passage
J. Huneke, G. Platero, and S. K\"ohler,
Phys. Rev. Lett. {\bf 110}, 036802 (2013).

\bibitem{Gurvitz}
%PHYSICAL REVIEW A 83, 042112 (2011)
%Quantum coherence and entanglement induced by the continuum between distant localized states
J. Ping, X.-Q. Li, and S. Gurvitz, Phys. Rev. A {\bf 83}, 042112 (2011).


\bibitem{Kayanuma-stoA}
Y. Kayanuma, J. Phys. Soc. Jpn. {\bf 53}, 108 (1984);
{\bf 53}, 118 (1984);
% Nonadiabatic transitons in level crossing with energy fluctuation. I. Analytical investigations.
% II. Numerical investigations
{\bf 54}, 2037 (1985).
% Sto theory for nonadiabatic level crossing with fluc off-diagonal coupling


%------------

\bibitem{Dykhne}
A. M. Dykhne,
%Adiabatic perturbation of discrete spectrum states
J. Exptl. Theoret. Phys. {\bf 41}, 1324 (1961) [Sov. Phys. JETP {\bf 14}, 941 (1962)];
L. D. Landau and E. M. Lifshitz,
{\it Quantum Mechanics (Non-relativistic Theory)}, 3rd Ed.,
Butterworth-Heinemann, Burlington (1977).

\bibitem{Wittig}
C. Wittig, J. Phys. Chem. B {\bf 109}, 8428 (2005).


\bibitem{OU}
G. E. Uhlenbeck and L. S. Ornstein, Phys. Rev. {\bf 36}, 823 (1930).


\bibitem{Makri}
N. Makri, J. Math. Phys. {\bf 36}, 2430 (1995).

\bibitem{MakriT}
K. Dong and N. Makri, 
%"Quantum stochastic resonance in the strong field limit", 
Phys. Rev. A {\bf 70}, 042101 (2004).

 	
\bibitem{KohlerE}
%Characterization of qubit dephasing by Landau-Zener interferometry
F. Forster, G. Petersen, S. Manus, P. H\"anggi, D. Schuh, W. Wegscheider, S. Kohler, and S. Ludwig,
arXiv:1309.5907. 

\end{thebibliography}
\end{document}